# From Individual to Social Imitation: How Communities Expand Organizational Search

**Esteve Almirall**
Esade Business School
esteve.almirall@esade.edu

**Christopher Tucci**
Said Business School, Oxford
c.tucci@imperial.ac.uk

## Abstract

Generative artificial intelligence illustrates a broader organizational puzzle: young, resource-constrained firms can build on knowledge produced across a wider field to address problems that exceed their internal experience. We theorize one mechanism as **social imitation**: a distributed and recursive process through which communities observe practices and outcomes across organizations, distill recurrent elements into portable strategies, circulate them, and revise the collective repertoire as adoption produces new evidence. This perspective endogenizes the object of imitation. Organizations do not merely copy practices; communities collectively produce what becomes available to copy.

We distinguish the source of a candidate—directly observed organizations or a collective—from its mode of adoption—blind or informed—and examine four forms of imitation in an agent-based model of organizational search. Collective sourcing alone does not improve performance. Its value depends on situated evaluation: assessing whether a candidate fits the adopter's configuration. As interdependence increases, informed social imitation improves population performance and the best solution discovered. As problems grow, it becomes more likely to outperform informed imitation from one or several directly observed peers because collective distillation sustains a broader candidate repertoire. Most gains arise from evaluating a small candidate set, and population performance can improve when only a minority of organizations possesses evaluative capability. Social imitation thus reveals a division of cognitive labor: communities expand the strategies organizations can consider, while organizations create value by determining which strategies fit. It explains how firms can mobilize knowledge beyond their boundaries—and why widespread access to collective knowledge does not produce equal benefits.

**Keywords:** imitation; organizational learning; vicarious learning; collective knowledge; organizational search; agent-based modeling; NK landscapes

## 1. Introduction

Generative artificial intelligence provides a contemporary illustration of the phenomenon that motivates this paper. Young firms can combine research, pretrained models, benchmarks, open-source tools, and engineering practices produced across a much larger field. The example does not establish that social imitation causes their speed or market performance, and the model below does not represent the AI industry. It makes visible a more general possibility: an organization can act on a repertoire of knowledge far larger than the experience accumulated within its own boundaries. That possibility has been studied as openness, as crowdsourcing, and as distant search (Almirall and Casadesus-Masanell 2010; Afuah and Tucci 2012; Bogers et al. 2017), and the level at which the relevant knowledge is produced has moved steadily above the individual firm.

Organizations rarely encounter managerial knowledge in the form in which it was first produced. Lean production, agile development, Six Sigma, continuous delivery, and professional standards are not complete replicas of one organization. They are portable abstractions assembled from observations across organizations, named and circulated by consultancies, professional associations, standards bodies, business schools, researchers, and the management press (Abrahamson 1996; Werr and Stjernberg 2003; David and Strang 2006). What reaches an adopter is therefore not simply another organization's behavior. It is a socially produced representation of collective experience.

Imitation theory has generated powerful explanations from a different starting point: an observing organization and a model it might copy. Organizations select salient or successful peers, infer which choices produced their performance, and reproduce some portion of those choices (Haunschild and Miner 1997; Rivkin 2000; Baum, Li, and Usher 2000). This dyadic view explains why imitation economizes on experimentation and why causal ambiguity makes complex practices difficult to transfer (Lippman and Rumelt 1982; Reed and DeFillippi 1990; Barney 1991). It also recognizes active imitators that choose whom and how much to copy (Csaszar and Siggelkow 2010; Nikolaeva 2014). Yet it leaves the production of many imitated objects outside the explanation. A field that distills experience from multiple organizations changes the source, form, and subsequent evolution of what organizations imitate.

We call this process **social imitation**: a distributed and recursive process through which a community observes practices and outcomes across multiple organizations, distills recurrent elements into portable abstractions, circulates those abstractions as candidates for adoption, and revises the collective repertoire as subsequent outcomes provide new evidence. Social imitation changes the unit of explanation from a copying relationship to a knowledge-production process. Organizations generate experience; communities and intermediaries transform it; adopters select and implement partial representations; and the results of adoption can enter later rounds of collective observation. No single actor need comprehend or control the whole cycle.

The change in unit of explanation raises a question that a dyadic theory cannot answer: when does collectively produced knowledge create value for an adopting organization? One plausible answer is circulation. A distilled practice reaches organizations that did not generate it and may embody regularities no single observer could identify. But portability does not imply local suitability. An abstraction omits much of the configurations in which it was observed, and under interdependence the same practice can have different consequences in different recipients (Szulanski 1996; Jensen and Szulanski 2004). Social imitation may therefore expand access without improving adoption.

We separate two dimensions that are often conflated. The **source** of a candidate may be individual, taken from one observed peer, or social, distilled from many organizations. The **mode of adoption** may be blind, applying a candidate without assessing its consequences locally, or informed, evaluating whether it fits before applying it. Crossing the dimensions yields blind and informed forms of individual and social imitation. This separation lets us ask three questions. Does a social source confer value by itself? When does situated evaluation improve social imitation? And if evaluation is also available to an individual imitator, when does the social source still matter?

We address these questions with an agent-based model of organizational search on NK landscapes. Firms first search locally. When local search is exhausted, they can copy a successful neighbor or draw practices from a catalog distilled from recurrent configurations among high performers. Adoption can occur from availability alone or after several candidates have been evaluated against the recipient's current configuration. The design holds spending, candidate size, repair, and information constant across the comparisons and adds informed individual arms that evaluate subsets of one peer or partial copies from several peers. We vary interdependence, problem size, evaluation breadth and quality, source-group composition, visibility of failure, and the share of firms able to evaluate.

Three findings define the contribution. First, **a social source alone is insufficient**. Blind social imitation never outperforms individual imitation because the adopter still cannot know whether a portable fragment fits its own configuration. Second, **situated evaluation creates local value**. Informed social imitation outperforms both blind forms as interdependence rises, provided that evaluation is at least moderately informative about local fit. However, an organization evaluating pieces of one peer captures essentially the same immediate gain. The social source does not make evaluation possible.

Third, **the source determines how long evaluation remains productive as problems grow**. One or several directly observed peers can supply only candidates tied to their configurations. Collective distillation instead transforms recurring combinations across many organizations into a broader repertoire of portable alternatives. Informed individual imitation is accordingly best in 31% of landscapes at the smallest problem size but only 8%, 6%, and 8% at the next three sizes. The pattern is a sharp decline to a small floor, not disappearance: direct peer observation sometimes contains enough useful variation, but it does not keep expanding the alternatives available to judgment.

The paper makes three contributions. The primary contribution is to **endogenize the object of imitation**. Rather than beginning with a practice already available to copy, social imitation explains how distributed organizational experience is selected, transformed, circulated, and revised as imitable knowledge. Second, we show that the social production of a candidate and the organizational creation of value are distinct. Communities can make experience portable, but adopters must establish situated fit. Third, we identify a source-dependent capacity constraint on selective imitation: evaluation is valuable only over the alternatives a source can supply, so collective distillation becomes consequential when problems outgrow direct observation of individual peers. These claims produce a conditional theory rather than a ranking. Social imitation is not generally superior; it is valuable when a collective can distill informative alternatives and organizations can evaluate them locally.

The model also bounds the argument. Evaluation has diminishing returns after roughly eight candidates on rugged landscapes, survives substantial noise, and fails when the signal is unrelated to prospective value. The collective must compare carriers with noncarriers, which requires observing unsuccessful organizations as well as winners (Denrell 2003). A minority of capable evaluators can improve population outcomes, but the benefit accrues to those firms rather than spilling over to blind adopters. These qualifications are not ancillary. They specify the organizational and field-level conditions under which social imitation can function as learning.

## 2. Social Imitation as an Organizational Process

### 2.1 From dyadic copying to collective knowledge production

Imitation research does not assume that organizations copy passively. Observers attend selectively, interpret others' experience, and choose an imitation breadth (Csaszar and Siggelkow 2010; Nikolaeva 2014). Firms imitate along different channels and for different reasons—by outcome, by frequency, by the status of the imitated (Haunschild and Miner 1997)—and they turn to others' experience when their own performance falls short of aspiration (Cyert and March 1963; Greve 1998; Gavetti et al. 2012). Vicarious learning may likewise involve active meaning-making and interaction rather than detached observation (Miner and Haunschild 1995; Myers 2018), and organizational heuristics can be ecologically rational responses to uncertainty (Gigerenzer, Reb, and Luan 2022). Our distinction is therefore not between passive prior imitators and active social imitators. It concerns how the object offered for imitation is produced.

In individual imitation, one organization is the source and another is the adopter. The observed configuration is incomplete evidence of the choices responsible for performance, making transfer vulnerable to causal ambiguity (Lippman and Rumelt 1982; Rivkin 2000). Ambiguity also obscures a firm's understanding of its own competences (King and Zeithaml 2001), so the exemplar cannot simply publish what makes it work, and replication and imitation come apart precisely at moderate complexity, where a firm can reproduce its own template but a rival cannot infer it (Rivkin 2001; Winter and Szulanski 2001). Imperfect observation can nevertheless preserve exploration by preventing exact convergence (Posen, Lee, and Yi 2013). Behind these results sits the routines tradition, in which what an organization knows is embedded in patterned activity rather than held as transferable propositions (Nelson and Winter 1982; Kogut and Zander 1992), articulated only through deliberate effort (Zollo and Winter 2002), and absorbed only by a recipient with prior related knowledge (Cohen and Levinthal 1990).

In social imitation, the candidate no longer comes directly from one model. Observations from multiple organizations are compared, decomposed, and transformed before they reach an adopter. The process distributes cognition across actors and separates the organizational origins of experience from its later representation and use. This is the level shift that openness, crowdsourcing, and distant-search research has been documenting from another direction: relevant knowledge is increasingly produced by a field rather than a firm, and the organizational question becomes how to draw on it (Almirall and Casadesus-Masanell 2010; Afuah and Tucci 2012; Bogers et al. 2017).

We define **social imitation** as a distributed and recursive process through which a community observes practices and outcomes across multiple organizations, distills recurrent elements into portable abstractions, circulates those abstractions as candidates for adoption, and revises the collective repertoire as subsequent outcomes supply new evidence. Six activities are analytically distinct: organizations experiment; practices and outcomes become observable; a community identifies recurrence across relevant cases; recurrent elements are distilled into portable abstractions; organizations select and adopt some abstractions; and adoption outcomes enter the evidence available for later distillation.

The process is social because observation, comparison, abstraction, circulation, and use are distributed across organizational and field-level actors. It is imitative because candidates originate in the observed choices of other organizations, even though no adopter copies a complete model. It is recursive because later outcomes may change the population that the community observes. Recursion does not guarantee improvement. Outcomes may validate, invalidate, or qualify an abstraction, but they may also be ignored, misread, or sampled selectively (Denrell and March 2001; Levinthal and Posen 2007).

Social imitation intersects with, but is not reducible to, adjacent conversations. Diffusion research explains how practices spread and how status, social pressure, and intermediaries shape adoption (Abrahamson 1991, 1996; Strang and Soule 1998; Abrahamson and Fairchild 1999; Rogers 2003). Social imitation makes the production and transformation of the circulating practice part of the explanation. Vicarious learning concerns learning from others' experience—including from their failures and near-failures (Baum and Ingram 1998; Kim and Miner 2007), and most strongly where uncertainty is highest (Srinivasan, Haunschild, and Grewal 2007); social imitation specifies a population-level architecture in which experience is aggregated and distilled before use. Research on managerial fashions and knowledge intermediaries examines actors that package and promote practices, showing that accounts of success are embellished in transmission (Zbaracki 1998), that a practice often spreads because it is ambiguous enough to be reinterpreted locally (Benders and van Veen 2001), that intermediaries proliferate and exit with belief rather than evidence (David and Strang 2006; Sturdy 2011), and that consultancies work precisely by abstracting cases into reusable methods (Werr and Stjernberg 2003). Social imitation treats those actors as possible carriers of a recursive process and distinguishes their production of candidates from an adopter's assessment of fit. A distilled practice may function as a socially produced heuristic, but social imitation denotes the process that produces, circulates, tests, and revises such heuristics.

These literatures can explain why organizations attend to, package, or adopt circulating practices. They do not ordinarily separate whether a candidate was produced from one organization or many from whether an adopter evaluates local fit. Nor do they predict how that separation interacts with problem scale. Making the architecture explicit generates the four forms below and the source-dependent capacity argument that distinguishes informed individual from informed social imitation.

The novelty of the construct therefore does not depend on the phrase itself. It lies in joining four elements usually studied separately: a population rather than one model as the source of experience; collective distillation rather than direct observation as the transformation mechanism; situated organizational evaluation as a condition of use; and adoption outcomes as inputs to later social knowledge. Figure 1 represents the process and the analytical distinctions used in the model.

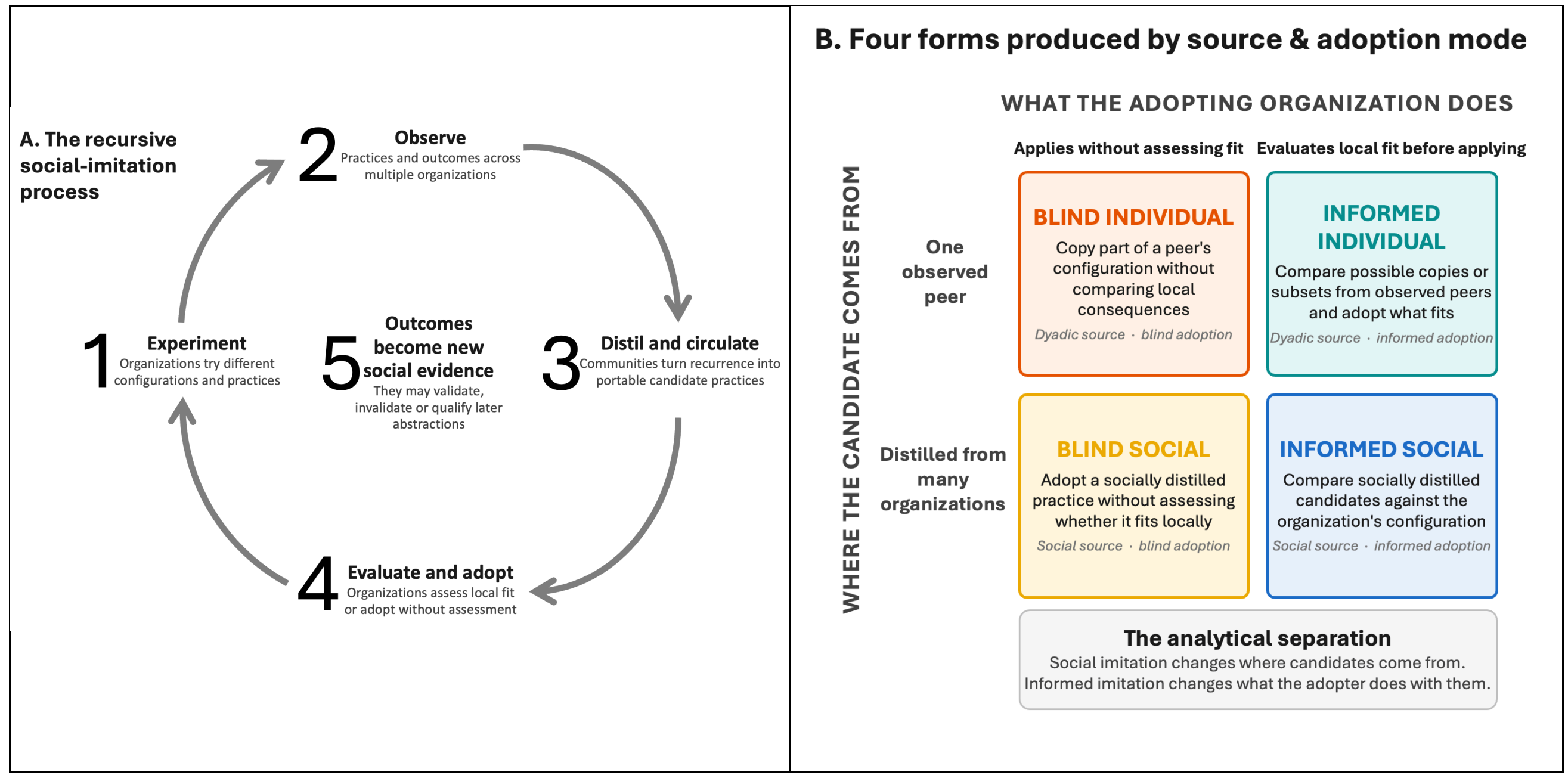

Figure 1. Social imitation as a recursive process and four forms of adoption. Panel A shows how organizational outcomes return evidence for later distillation. Panel B separates the source of a candidate from whether the adopter evaluates local fit.

## 2.2 Source and adoption mode

Two dimensions distinguish four forms of imitation. The first is the **source of the candidate**: one observed peer or patterns distilled from multiple organizations. The second is the **mode of adoption**: applying a candidate without assessing local consequences or evaluating whether it fits before applying it. The second dimension separates two activities that organizational search research treats as distinct with distinct returns—generating alternatives and evaluating them (Knudsen and Levinthal 2007)—and the architecture through which an organization evaluates is itself consequential for what survives screening (Sah and Stiglitz 1986; Csaszar 2013; Csaszar and Eggers 2013).

**Blind individual imitation** copies part of one peer's configuration without comparing alternatives for local fit. **Informed individual imitation** generates possible copies or subsets from one or more peers, evaluates them against the recipient's configuration, and adopts what fits best. **Blind social imitation** adopts a collectively distilled practice without assessing its local fit. **Informed social imitation** considers several socially distilled practices and adopts a candidate only when it is expected to improve the recipient's configuration.

This typology prevents two errors. Social imitation is not defined by superior performance: blind and informed variants belong to the same production process even if one fails. Evaluation is also not unique to social imitation: an organization can deliberate over candidates from a peer. The theoretical question is how source and adoption mode interact as problems change.

### 2.3 Collective source without situated evaluation

Collective distillation makes organizational experience portable without making it appropriate for every recipient. A distilled practice identifies recurrence while omitting much of the configurations in which that recurrence appeared. When organizational choices are interdependent, the same bundle can have different consequences alongside different remaining choices (Simon 1962; Rivkin and Siggelkow 2007). Recurrence among successful organizations is therefore evidence that a practice merits consideration, not that it will improve every adopter—and what an organization can attend to is itself bounded (Ocasio 1997).

Circulation increases availability but does not resolve fit. Blind social imitation can reproduce the central problem of individual imitation in a different form: the adopter receives a decomposed object but cannot infer its consequences locally. This yields a deliberately restrictive expectation:

**Hypothesis 1 (H1). Blind social imitation will not outperform individual imitation solely because it draws on collectively distilled practices.**

H1 identifies what collective production cannot accomplish alone. If blind social imitation performs no better than peer copying, any advantage of its informed variant must arise from how candidates are used rather than simply from access, bundle size, expenditure, or source.

### 2.4 Situated evaluation under interdependence

Situated evaluation compares candidates against the recipient's existing configuration. It asks a prospective rather than retrospective question: not whether a practice accompanied success elsewhere, but whether it is likely to improve this organization given the choices already in place. In this respect it is closer to forward-looking cognitive search than to backward-looking experiential search (Gavetti and Levinthal 2000), and the practice functions as an analogy applied to the adopter's own position rather than a template to be reproduced (Gavetti, Levinthal, and Rivkin 2005). The capability is fallible. Its value depends on whether the signal used to rank candidates is informative about local fit.

The return to evaluation should rise with interdependence. When choices are nearly separable, importing a successful peer's choice or an isolated practice produces few damaging interactions, and local search can correct errors one choice at a time. As interdependence rises, the same practice can have different consequences in different recipients (Levinthal 1997; Rivkin 2000; Fleming and Sorenson 2001). External success becomes less informative about local value, and comparing alternatives before commitment becomes more useful.

**Hypothesis 2 (H2). Provided that evaluation is informative about local fit, the advantage of informed social imitation over blind social and individual imitation will increase with interdependence.**

The qualification makes signal quality a boundary condition rather than a hidden assumption. Breadth may compensate for imprecise evaluation, but it cannot rescue a signal unrelated to prospective value.

### 2.5 Why the social source matters as problems grow

If evaluation produces the immediate gain, an organization might obtain it by considering pieces of one peer's configuration or by evaluating partial copies from several peers. Informed individual imitation should indeed be competitive when directly observed peers contain sufficient useful variation. The distinctive contribution of collective distillation is not that it makes evaluation possible, but that it transforms observations across organizations and changes the candidate supply over which evaluation operates.

Let an adopter hold configuration $x$ and observe peer $y$. Every candidate drawn from that peer assigns some subset of choices the values that $y$ holds. Considering more subsets can approach the best recombination available from $y$, but it cannot supply a value absent from that configuration. Observing additional peers increases the number of direct candidates but leaves each candidate attached to a particular observed configuration. A socially distilled catalog instead identifies agreements across different subsets of high-performing organizations. It can contain alternative values and combinations supported by different cases. Attention and support thresholds still bound the catalog, but the resulting candidates are products of cross-case transformation rather than copies tied to one model.

The difference should become consequential as problem size grows. Larger problems create more possible locations of error and more combinations an adopter may need to consider. A limited set of peers remains a bounded collection of points in that expanding space, whereas collective distillation continues to draw on population-level variation. The same logic underlies the argument that broadcasting a problem to a wider field substitutes for a firm's own costly local search (Afuah and Tucci 2012), and the observation that openness widens the discoveries available while pulling designs away from any one organization's trajectory (Almirall and Casadesus-Masanell 2010).

**Hypothesis 3 (H3). As problem size grows, informed social imitation will become more likely to outperform informed individual imitation because collective distillation sustains a broader candidate supply than direct observation of individual peers.**

H3 predicts a change in relative advantage, not universal dominance. Informed individual imitation may remain best in some landscapes, and the model caps the catalog. The relevant pattern is a reduction in the regime where peer evaluation is best, not its necessary disappearance.

## 3. Model

### 3.1 The setting in words

The model puts a hundred firms in the same industry and gives each of them the same kind of problem: a set of choices that interact, so that whether any one choice is right depends on the others already in place. Each firm starts somewhere at random and improves what it can on its own, changing one choice at a time and keeping whatever helps. Sooner or later that runs out. Every remaining single change makes things worse, even though better arrangements exist elsewhere. This is the moment the paper is about: a firm has exhausted what it can work out alone, and must decide what to take from outside.

It has two kinds of option. It can look at the ten firms it can observe, pick the one performing best, and copy part of what that firm does—imperfectly, because it cannot see everything. Or it can turn to a catalog. Each period a community looks across the whole population, identifies the combinations of choices that recur among the ten best performers, and publishes them as small named bundles. Any firm can take one.

What the firm then does with what it has is the second decision, and it is the one that matters. It can apply the candidate as it arrives. Or it can hold several candidates against its own current arrangement, work out what each would be worth *here*, and adopt only the best of them, and only if it beats standing still. Adoption is not costless and not clean: importing a bundle leaves the rest of the organization mismatched, so a few local repairs follow, and they are charged to the same allowance as everything else. Every firm gets the same allowance of moves, so no arrangement can win by simply doing more.

Because the community re-reads the population every period, what firms adopt today shapes what is available to distill tomorrow. The catalog is an output of the population as well as an input to it.

### 3.2 Landscape, population, and timing

We use an agent-based model because the argument concerns the interaction of organizational search, population-level distillation, and repeated adoption—a set of interdependent processes that is difficult to characterize analytically and for which simulation is the appropriate instrument (Davis, Eisenhardt, and Bingham 2007; Harrison et al. 2007; Burton and Obel 2011; Fioretti 2013). Firms search an NK landscape (Kauffman 1993; Levinthal 1997; Csaszar 2018), a standard representation of organizational choice under interdependence (Ganco and Hoetker 2009; Baumann, Schmidt, and Stieglitz 2019). A configuration is a binary string of *N* choices. Each choice contributes a value conditional on its own setting and the settings of *K* other choices; fitness is the mean contribution. Increasing *K* increases interdependence.

Formally, firm *i* holds configuration $x_i$ in $\{0,1\}^N$. Its fitness is the mean of *N* choice contributions:

$$F(x_i) = N^{-1} \sum_{n=1}^{N} f_n(x_{in}, x_i, V(n))$$

Here *V(n)* contains the *K* choices interacting with choice *n*. At round *t*, the community forms catalog $C_t$ from closed bundles *a* held by at least round($E \cdot \tau$) of the *E* exemplary firms and whose size is between 2 and *N*/3. At $E = 10$ and $\tau = 0.25$ that support floor is two firms. Applying *a* to a firm's configuration, written $x_i \oplus a$, replaces the values of the choices specified by the bundle.

We examine $N = 32$, 48, and 64 across uniform grids from $K = 0$ to $N - 1$, using 500, 200, and 100 replications per cell. A fourth size, $N = 96$, uses 13 interdependence levels and 50 replications. Fitness is min-max normalized within each landscape against an estimated floor and ceiling. The Online Appendix gives the estimator, raw-unit conversions, full cell-level results, and a replication of a published NK imitation model.

One hundred firms occupy a Watts-Strogatz ring of degree 10 and rewiring probability 0.1. Each observes ten peers and their realized fitness. Firms begin at random configurations and act for 200 rounds. In each round, a firm first makes the best available one-choice improvement. Only when no local change helps does it act socially. The model therefore compares forms of imitation after local search has been exhausted.

### 3.3 Individual and social imitation

The principal arms implement the 2 × 2 typology. **Solitary search** stops at a local optimum. **Blind individual imitation** identifies the best observed neighbor, copies each choice with probability 0.30, and repairs locally. The copy probability represents partial observability and sits near the benchmark's own optimum; a robustness analysis varies it through perfect copying.

For social imitation, a community observes the population each round and identifies closed patterns that recur among the top ten firms. A pattern must be held by at least two of that group, is limited to *N*/3 choices, and the hundred best-supported patterns are published. The mining procedure is exhaustive over the intersection lattice of elite configurations; the Online Appendix proves that this enumerates every closed pattern exactly at a cost that does not grow with problem size. The catalog is re-created each round, so adoption changes the population from which later practices are distilled. The model thereby represents the performance-learning core of recursion while omitting legitimacy, persuasion, strategic reporting, competing intermediaries, and an accumulated historical stock of named practices. Because the reported design does not compare this updating catalog with a fixed one, it does not identify the incremental value of recursion itself; the hypotheses concern the source and use of candidates within the recursive architecture.

**Blind social imitation** receives and applies one available candidate without ranking local alternatives. **Informed social imitation** draws *m* candidates, estimates the value each would have in the firm's current configuration, and adopts the best only if it improves on the status quo. The primary specification uses *m* = 5 and perfect prospective evaluation; additional runs vary breadth, error, and the correlation between the evaluation signal and prospective value.

For a sampled set $S_{it}$ contained in $C_t$, informed adoption follows the rule:

$$a^* = \arg\max_{a \in S_{it}} \hat{F}_i(x_i \oplus a);\ \text{adopt iff } \hat{F}_i(x_i \oplus a^*) > F(x_i)$$

Under prospective evaluation, *F-hat* equals the landscape value of the modified configuration; robustness treatments add error or replace it with a signal correlated with prospective value. This rule isolates the model's central distinction: the catalog determines what can be considered, whereas the adopter's evaluation determines which candidate, if any, enters its configuration.

To isolate source from evaluation, **informed individual imitation** uses the same evaluation, acceptance, repair, and accounting rules but draws candidate subsets from one peer's configuration. A second version evaluates partial copies from several peers. A size-matched random-bundle placebo tests whether population mean fitness can be raised through homogenization rather than effective search.

### 3.4 Cost, comparability, and research sequence

Each firm receives a budget of 20 moves. A social action, a local-search step, and each postadoption repair step cost one move; repair is capped at five. Repair is not a technical convenience: importing a bundle leaves the remaining choices tuned to a configuration that no longer exists, and recipients of transferred practice are known to adapt what they receive rather than install it unchanged (Szulanski 1996, 2000; Winter and Szulanski 2001; Jensen and Szulanski 2004). The cap represents bounded attention, and repair may not touch the choices the social move has just set, so it adapts an adoption rather than undoing it. Pattern mining and evaluation are not charged in the primary model. The interpretation is therefore the value of deliberative capability, not its net economic return. A charged-evaluation robustness test leaves the result unchanged at plausible costs, and a separate specification denominates the budget in changed choices rather than moves.

Every comparison reports spending, choice-set size, and choices changed per move. Contrasts are treated as interpretable only when spending differs by no more than 10%. The reported main comparisons fall between 0.947 and 1.039, with two declared exceptions flagged where they occur in the Online Appendix. Social arms can change more choices per move, but this does not explain the findings: blind social imitation changes the most at the first three sizes and gains nothing; at $N = 96$ the asymmetry reverses and informed social imitation still wins; and the advantage survives when budgets are denominated in changed choices.

The hypotheses organize the final theoretical comparisons; they should not be read as a chronological account of model development. The Online Appendix documents the iterative sequence that affected inference. An initial sampling miner was replaced by an exhaustive procedure so problem size would not be confounded with algorithmic sampling; two claims that failed after the change were removed. Four earlier comparator verdicts were vacated when cost or choice-set inequalities were found. We report the affected designs, failed expectations, complete robustness grid, and an independent implementation of a published model. This disclosure follows calls to make simulation assumptions and bounded-rationality choices explicit (Burton and Obel 2011; Puranam et al. 2015). Table 1 summarizes the parameters used in the reported runs.

**Table 1. Principal Model Parameters**

| Parameter | Primary value | Variation |
|---|---|---|
| Problem size ($N$) | 32, 48, 64, 96 | Main treatment |
| Interdependence ($K$) | 0 to $N-1$ | Main treatment |
| Firms / observed peers | 100 / 10 | 250 and 500 firms |
| Rounds / move budget | 200 / 20 | Budget 24 |
| Replications | 500 / 200 / 100 / 50 | By $N$ |
| Copy probability | 0.30 | 0.10 to 1.00 |
| Exemplary group | Top 10 firms | Count versus share |
| Support floor / length / catalog cap | 2 of 10 / $N$/3 / 100 | Mining robustness |
| Candidates evaluated ($m$) | 5 | 1, 2, 3, 5, 8, 10 |
| Evaluation error | 0 | 0 to 64 times candidate spread |

## 4. Results

### 4.1 A social source alone does not create value

H1 compares blind social imitation with blind individual imitation. Both operate at matched spending and receive similarly sized adopted bundles. Blind social imitation also uses the same catalog and candidate supply later used by the successful informed arm. It differs only in not comparing candidates for local fit.

The social source alone produces no advantage. On the best solution found, its contrast with individual imitation is −0.001, +0.005, and +0.000 at $N = 32$, 48, and 64; every interval includes zero. At high interdependence it is slightly worse on population mean fitness. Across all 69 landscapes, blind social imitation is never the uniquely best form. These results support H1. Collective distillation creates an object that can circulate, but circulation alone does not convert it into organizational value.

**4.2 Situated evaluation becomes more valuable under interdependence**

H2 compares informed social imitation with both blind forms. Figure 2 reports the central pattern. Near $K = 0$, the arms converge because choices can be corrected independently. As interdependence rises, informed social imitation separates from both individual and blind social imitation on population mean fitness, best solution found, and retained diversity.

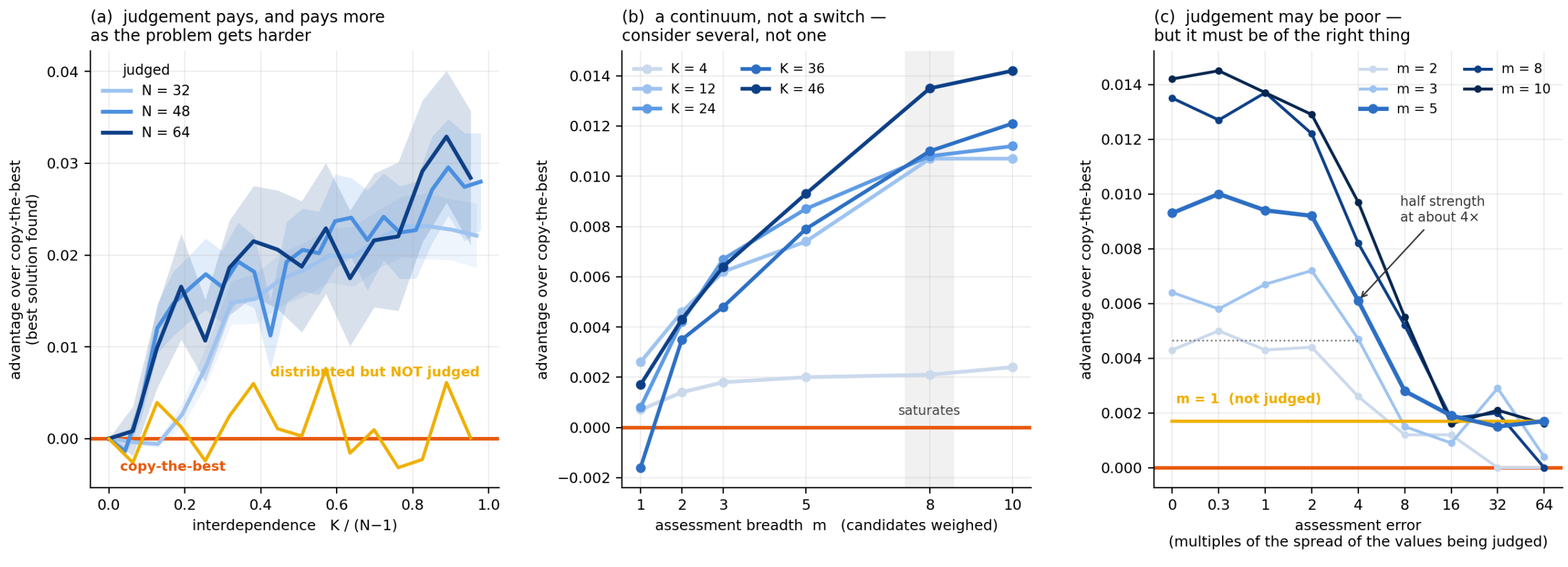

Figure 2. Situated evaluation and its limits. Panel (a) shows the advantage of informed social imitation over individual imitation as interdependence rises, with blind social imitation near zero. Panel (b) shows diminishing returns to evaluation breadth. Panel (c) shows robustness to evaluation error. N = 48 in panels (b) and (c); matched spending; 95% Student-t intervals.

At the hardest tested interdependence, informed social imitation exceeds individual imitation on best solution found by +0.022, +0.028, +0.028, and +0.028 at $N = 32$, 48, 64, and 96. Expressed relative to the gain that individual imitation itself provides over solitary search, evaluation adds 42%, 58%, 59%, and 93%. The absolute increment does not grow after $N = 48$; its relative importance rises because the value of imitation without situated evaluation declines as problems grow. On population mean fitness, evaluation adds +0.006, +0.020, +0.029, and +0.037, equivalent to 20%, 94%, 164%, and 252% of the corresponding imitation gain.

The interdependence at which informed social imitation begins to win also falls with problem size. At the primary budget, the threshold is $K \geq 14$, 8, and 4 for $N = 32$, 48, and 64; the second budget reproduces 14 and 8 but moves the $N = 64$ threshold to 8. We therefore use the conservative sequence 14/8/8. At $N = 96$, the arm wins 12 of 13 tested levels beginning at $K = 8$. These results support H2 while locating its force in the persistence of evaluation's value rather than a growing absolute effect.

Evaluation is graded and bounded. At $N = 48$ and high interdependence, the advantage over individual imitation rises from +0.005 with one candidate to +0.012, +0.018, +0.026, +0.038, and +0.040 with 2, 3, 5, 8, and 10 candidates. The curve saturates at approximately eight candidates on rugged landscapes and three on smoother ones.

The result is also robust to imperfect judgment within limits. With five candidates, the advantage remains +0.017 when evaluation error is four times the dispersion of the values being ranked and falls to +0.008 at eight times that dispersion. Considering more candidates shifts the decay curve outward. Breadth therefore substitutes for noisy evaluation over a range. It does not substitute for an irrelevant signal. When the ranking signal is correlated with prospective value at $\rho = 0$, 0.1, 0.2, 0.3, 0.5, 0.7, and 1.0, the advantage is +0.002, +0.004, +0.009, +0.011, +0.015, +0.023, and +0.029. Roughly half the available gain returns near $\rho = 0.5$; a signal largely unrelated to local fit produces none. H2's informativeness condition is therefore substantive.

Perfect copying does not overturn the result. When the individual imitator observes and transplants its best peer's complete configuration, it collapses the population toward one solution and performs worse on the frontier. Its own best-found performance peaks at partial fidelity of 0.10–0.20, with the primary value of 0.30 statistically indistinguishable from that peak. Imperfect imitation is a source of exploration rather than simply a defect (Posen, Lee, and Yi 2013), and the same logic by which restricted sampling distorts what a population learns applies here (Denrell and March 2001).

### 4.3 Collective sourcing matters as problems grow

H3 uses two demanding informed-individual comparators. One evaluates subsets of the best peer's configuration; the other evaluates a partial copy from each of several high-performing peers. Both use the same evaluation function, acceptance rule, repair, and cost accounting as informed social imitation. The single-peer subset arm captures essentially the entire immediate benefit of evaluation. At the hardest interdependence, it improves on blind individual imitation by +0.023, +0.028, and +0.031 on best solution found at $N = 32$, 48, and 64, compared with total informed-social gaps of +0.022, +0.028, and +0.028. Situated evaluation, not collective sourcing, creates the immediate frontier gain.

The sources diverge as problem size increases. The single-peer subset arm has only 2.5–3.8 distinct candidates available after duplicates, compared with 27–41 from the catalog. Directly sampling several peers raises variety, but it does not remove the capacity pattern. Across the 56 cells of the three dense grids, informed social imitation is best in 40, one of the two peer-evaluating arms in 8, blind individual imitation in 3, and no form is separable in 5. The peer-evaluating category is best in five of sixteen landscapes at $N = 32$ but only two of twenty-four at $N = 48$ and one landscape at each of $N = 64$ and 96. Table 2 reports whichever informed-individual comparator performs best in each cell. Its competitive share falls sharply from 31% to 8%, 6%, and 8%, consistent with a source-dependent capacity constraint and supporting H3.

**Table 2. Best-Performing Form by Problem Size**

| *N* | Landscapes | Informed social | Informed individual | Blind individual | Not separable | Informed-individual share |
|---|---|---|---|---|---|---|
| 32 | 16 | 7 | 5 | 2 | 2 | 31% |
| 48 | 24 | 19 | 2 | 1 | 2 | 8% |
| 64 | 16 | 14 | 1 | 0 | 1 | 6% |
| 96 | 13 | 11 | 1 | 0 | 1 | 8% |
| **All** | **69** | **51** | **9** | **3** | **6** | **13%** |

The $N = 96$ run uses a thinner grid, fewer replications, and five rather than six arms. We therefore treat its single informed-individual win as evidence that the decline does not reverse, not as a precise estimate of an 8% landscape share or evidence of a rise from 6%.

Figure 3 maps the regimes on the three dense grids. When choices are separable, form makes little difference. At low interdependence, copying a peer is best. A band of smaller, moderately interdependent problems favors informed evaluation of directly observed peers. At high interdependence, informed social imitation dominates most cells, and its threshold falls with problem size. Blind social imitation occupies no regime.

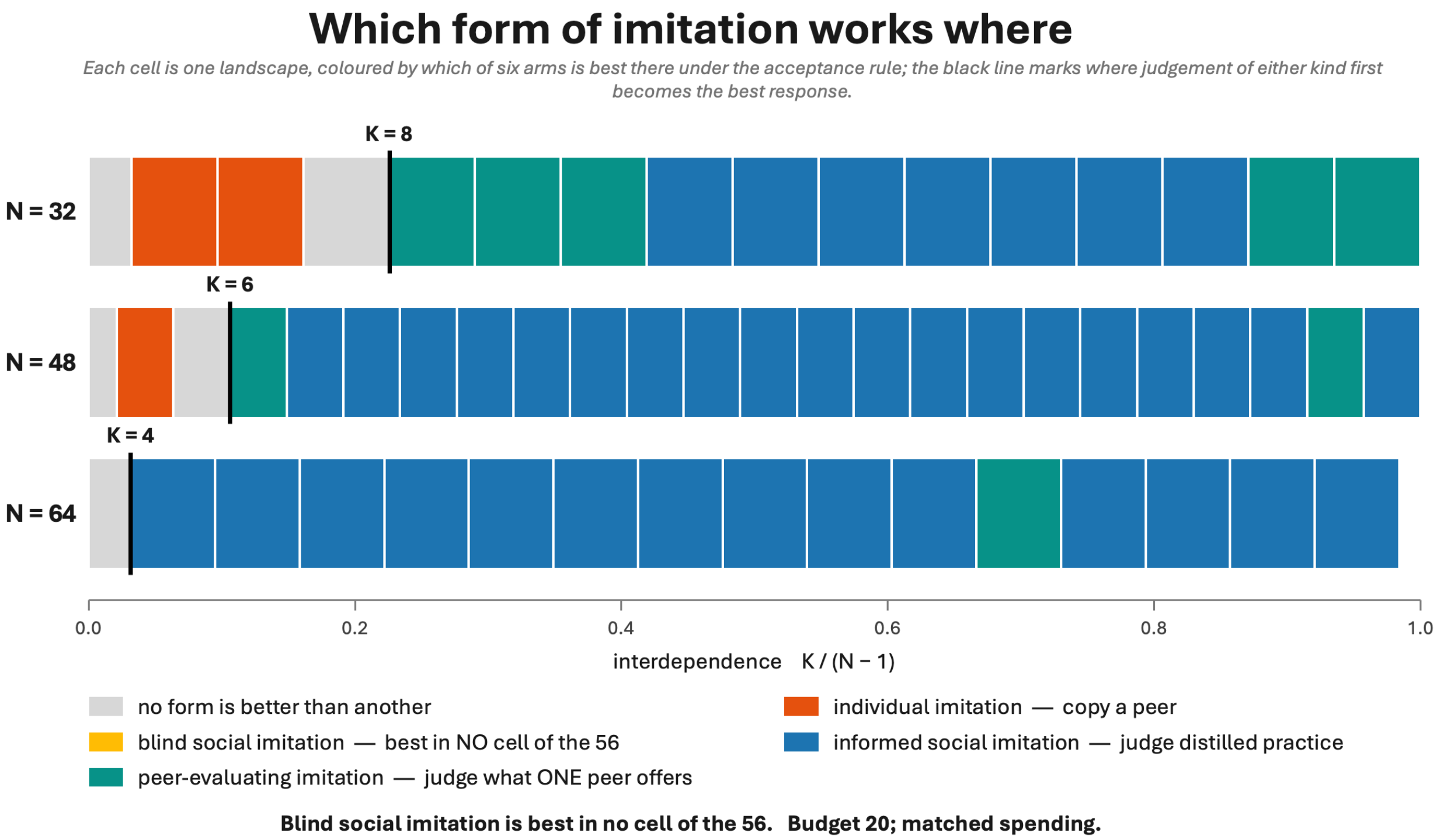

Figure 3. Which form of imitation works where. Each cell is one landscape from the three dense grids, classified by the best-performing arm under the prereported acceptance rule; the line marks where informed adoption first becomes the best response. Blind social imitation is best in no cell. Budget 20; matched spending.

### 4.4 Conditions on collective production and organizational capability

The social process must produce an informative catalog. Catalogue size alone is not evidence: a catalog mined from ten random firms contains 95 patterns, compared with 80–82 from the ten best. Rarity-based scoring also fails because any small group sharing a bundle produces high within-group support relative to the population. What separates informative from coincidental recurrence is the fitness difference between carriers and noncarriers. This criterion removes 47% of a random catalog but 17% of the real catalog and leaves the H2 winning region unchanged.

The comparison works only when the carrier-noncarrier difference is estimated across the population. Within the uniformly high-performing exemplary group, there is too little outcome variance to determine which shared practices contributed to performance. A collective must therefore observe unsuccessful organizations as well as winners. This extends Denrell's (2003) undersampling argument: missing failure does not merely bias estimates of a practice's value; it can eliminate the variation needed to identify value at all. It is also why learning processes biased toward the near, the recent, and the successful are a hazard at the level of the collective and not only of the firm (Levinthal and March 1993; Denrell 2005).

Selectivity also matters more than population scale. Increasing the number of firms from 100 to 250 and 500 does not enlarge the advantage of informed social imitation, while individual imitation benefits from better available neighbors. Keeping the exemplary set at ten firms works at each population size. Defining it instead as 10% of the population enlarges the group until no sufficiently specific configuration recurs often enough, emptying the catalog. Collective distillation requires a sharply defined source, not simply more observations.

Finally, not every organization must evaluate. In heterogeneous populations, the smallest share of informed adopters that improves best-found performance over an all-blind population ranges from 5% to 50% across six focal cells and is 10% or 25% in four. Aggregate returns are approximately linear in the judging share, with neither a threshold nor early saturation. The gains are not public, however. Mean performance among nonjudging firms is unchanged or slightly lower as the judging share rises, and the catalog itself barely changes. Population performance improves because capable firms improve, not because they lift blind adopters. Who evaluates matters less consistently than how many: assigning the capability to initially high-performing firms produces two null differences and one positive difference across the three sizes.

Together, the results identify a conjunction rather than a generally superior architecture: social imitation creates value when problems are sufficiently interdependent, the collective distills a selective and outcome-informative candidate set, and adopters possess a signal that bears on local fit. Remove situated evaluation and the social source creates no advantage; restrict the source to one configuration and evaluation encounters a capacity constraint as problems grow.

## 5. Discussion

### 5.1 Imitation as a socially organized process

Imitation theory commonly locates learning in a relationship between an observer and a model. Our first contribution is to move the production of the imitated object inside the theory. Social imitation begins before an adopter encounters a practice and continues after adoption. Multiple organizations generate heterogeneous experience; communities and intermediaries select, compare, and abstract it; adopters interpret and implement portable fragments; and later outcomes can alter the evidence available for subsequent distillation. This distributed architecture is not individual imitation with more models. It changes who performs the cognitive work, how experience is transformed, and what later becomes available to imitate.

The construct also clarifies the relation between imitation and diffusion. Diffusion explains the movement and uptake of practices, including the roles of status, legitimacy, and social pressure. Social imitation asks where the circulating practice came from and how collective experience was transformed into that object. Its recursive element links adoption back to later knowledge production. The model isolates only the performance-learning core of this broader process, but doing so reveals that neither social production nor circulation guarantees improvement.

### 5.2 Separating social production from situated value creation

The second contribution is a separation between producing a candidate and creating value from it. Collective distillation identifies recurrence and makes partial knowledge portable. It cannot establish how an abstraction interacts with the unrepresented remainder of every recipient's configuration. That judgment remains situated. Blind social imitation consequently performs no better than peer copying, while informed adoption performs substantially better as interdependence rises.

This finding should not be read as relocating the paper's contribution from social imitation to judgment. Evaluation is already recognized in theories of selective imitation and organizational decision making (Csaszar and Siggelkow 2010; Knudsen and Levinthal 2007). Its role here is to expose what a social process can and cannot do. Collective production expands the domain of possible action; organizations still create local value by assessing fit. The distinction explains why widely circulated best practices can coexist with heterogeneous outcomes and why better dissemination alone may fail to improve adoption.

The heterogeneous-population result sharpens the implication. A minority of informed adopters can improve the population, but blind firms do not share the gain. Social imitation is therefore not automatically a public good. Collective availability and private absorptive capability are complements (Cohen and Levinthal 1990), and the return can remain concentrated in organizations able to perform the situated work.

**5.3 A source-dependent capacity constraint**

The third contribution identifies when the social source becomes theoretically consequential. Selective imitation is limited by both evaluation and candidate supply. One peer can be decomposed and assessed, and candidates from several peers can be compared, so collective sourcing does not uniquely produce the immediate benefit of deliberation. Yet direct observation remains tied to a limited set of configurations. As problems grow, population-level distillation can sustain alternatives produced from recurrence across cases rather than supplied by any one observed model.

This capacity constraint generalizes beyond the particular catalog in the model. An organization learning from one alliance partner, acquired unit, benchmarking visit, or exemplar can deliberate only over what that source provides. A consortium, multi-client study, professional association, or standards process can combine recurrence across sources (Afuah and Tucci 2012; Almirall and Wareham 2011). The theory predicts that the difference should widen with the size and interdependence of the practice domain, not simply with the prestige or average quality of either source.

The constraint also changes the role of intermediaries. Their most useful output is not a universal recommendation but an interpretable candidate set. The model suggests that candidate sets should be selective, include evidence from failures, and preserve alternatives for local comparison. Because what circulates is a representation rather than the organization it came from, its usefulness depends on how well that representation travels to configurations unlike the ones it was drawn from (Massa, Tucci, and Afuah 2017; Chesbrough and Tucci 2020). The diminishing return to breadth is encouraging: informed social imitation does not require unbounded attention. A handful of seriously evaluated candidates captures most of the modeled gain. This is an attention-efficiency result, not an estimate of the full economic cost of producing and evaluating collective knowledge.

Generative artificial intelligence illustrates how this mechanism might operate without constituting evidence for it. A young firm may act as a selective integrator of research, models, benchmarks, tools, and practices generated across a wider technical community. Social imitation would predict advantage only when that field-level repertoire is informative and the firm can evaluate and recombine candidates for its own architecture and market—which is a question about how external knowledge is integrated with what an organization already has, not merely about access to it (Lanzolla, Pesce, and Tucci 2021). The model does not show that social imitation makes AI firms faster or more successful; testing that application would require evidence on firms' knowledge sources, evaluative routines, time to performance, and outcomes.

### 5.4 Boundary conditions, limitations, and future research

The results define four principal boundary conditions. First, evaluation matters only when choices interact; on smooth problems, copying a good peer is already effective. Second, the evaluation signal must be informative about prospective fit. Breadth offsets noise but not irrelevance. Third, collective distillation requires outcome variation, including unsuccessful cases, and a selective exemplary group. Fourth, social sourcing becomes distinctive with problem scale; informed individual imitation remains competitive on some small and some large landscapes.

The model makes strong abstractions. Evaluation is uncharged in the primary specification, although light charges and choice-denominated budgets preserve the result. Firms evaluate prospective value more directly than real organizations typically can; the signal experiments show how performance deteriorates as that informational advantage is removed but do not explain how organizations construct valid signals. The catalog is re-mined rather than historically accumulated, so the model cannot represent abstractions built on prior abstractions or path-dependent learning sequences (Bingham and Davis 2012). Re-mining allows adoption outcomes to affect later catalogs, but without a fixed-catalog comparison the model does not estimate the incremental contribution of that recursion. The environment is static; where landscapes shift, the returns to continued search change and any fixed distilled practice decays with them (Posen and Levinthal 2012). Firms have no internal structure, so the model cannot represent search that improves one organizational level while degrading another (Siggelkow and Rivkin 2005, 2006). Results come from one landscape family and network topology, and near-decomposable structures produce no additional advantage (Simon 1962; Ethiraj and Levinthal 2004). The $N = 96$ analysis uses fewer interdependence levels and replications than the three primary sizes, so it establishes persistence of the scale pattern rather than its behavior beyond that range.

These limitations define an empirical agenda. The theory predicts that organizations in more interdependent domains will invest more in pilots, diagnostic work, and comparison before adoption; that the value of practice-distilling intermediaries will rise with domain size and interdependence; and that deliberation over one source will saturate sooner than deliberation over collectively generated alternatives. Field research can examine how intermediaries choose cases, whether failed adoptions enter their evidence, how adopters evaluate local fit, and whether adoption outcomes actually update later practice repertoires. Models with competing distillers, strategic reporting, moving landscapes, and persistent catalogs can test the recursive dynamics omitted here.

## 6. Conclusion

Organizations imitate more than other organizations. They imitate practices that communities have produced from the experience of many. Treating this as social imitation changes the theory from a dyadic account of copying to a distributed account of how imitable knowledge is produced, transformed, used, and renewed.

The social process is valuable, but not automatically. Collective distillation without situated evaluation creates no advantage. Evaluation creates local value wherever candidates originate, while collective sourcing becomes consequential when problems outgrow what directly observed peers can supply. Social imitation therefore rests on a division of cognitive labor: communities expand and refresh the alternatives available for consideration; organizations determine which, if any, belongs in their own configuration.

Two implications of that division deserve emphasis, because both concern conditions that no single organization can secure for itself.

**The collective side depends on an open and varied public discussion of practice.** Our results make two demands of the distillation step that are ordinarily treated as matters of taste. The first is that failure has to be visible. A collective that observes only winners has too little outcome variance to tell which shared practices contributed to performance, so the distillation is not merely biased but uninformative in principle—a stronger conclusion than undersampling alone implies. Organizations publicize adoptions that worked and rarely those that did not, and the practice literature does much the same; researchers are among the few actors with an incentive to record what did not pay, which makes that record a precondition for the mechanism rather than a marginal contribution to it. The second demand is variety. The catalog exists only because organizations differ: recurrence carries information precisely against a background of configurations that do not share it. Anything that drives a population toward a common configuration erodes the raw material of distillation, and our own model shows the mechanism at work—the arrangements that homogenize the population, whether random adoption or perfect copying, register well on average performance while finding worse solutions and retaining fewer distinct ones. A field in which everybody has converged on the same practices has nothing left to distill. A healthy diversity of organizational arrangements, and a discussion in which unsuccessful adoptions can be reported without penalty, are not decoration around the process. They are its inputs.

**The organizational side depends on intermediaries and firms that understand their domain and choose well.** The model is unambiguous that circulation is not the binding constraint. What binds is whether the candidate set is informative and whether the adopter can tell which member of it belongs in its own configuration. For intermediaries—consultancies, associations, standards bodies, the practice literature—this reframes the product. A universal recommendation is the $m = 1$ condition of our design, which is the control arm that gains nothing. The useful output is a small, separable, honestly evidenced set of alternatives that an organization can hold against its own arrangement, and about eight of them exhausts the available gain on hard problems. That is a demanding standard in a different way from the usual one: it requires enough domain knowledge to know which alternatives are worth putting in front of a client, and enough restraint not to publish more. For adopting organizations, the requirement is a judgment that bears on their own situation rather than on the reputation of the source. Our signal experiments put a number on how good that judgment must be—a correlation with prospective value of roughly one half recovers half the available gain, while ranking practices on what they publicly carry recovers essentially none—and our heterogeneity results show that a minority of organizations with that capability can lift a population's performance, though the gain accrues to them rather than to the firms that adopt blindly. Widely available knowledge and privately held judgment are complements, not substitutes, and the returns to collective knowledge production are captured by those who can tell what fits.

That distinction explains both the promise and the limits of learning from collectively distilled practice. The promise is that an organization can act on far more experience than it could ever accumulate. The limit is that no amount of collective experience decides what belongs in any particular organization.

## Declarations

**Data and code availability.** The model, mining code, experimental scripts, frozen empirical dataset, and cell-level result tables are archived and will be made available to editors and reviewers through the submission system. The Online Appendix reports an independent replication of a published model using the same code base.

**Funding.** No funding involved in this project.

**Competing interests.** The authors declare no competing interests.

**Use of AI-assisted technologies.** Large language models were used for assisting in code implementation, and editorial work on the manuscript. The authors designed the model and experiments, determined the claims, and verified every reported result against archived output.

# Online Appendix for "Improving Today, Narrowing Tomorrow: Collective Learning, Diversity, and Generativity"

## A. Formal specification

### *A.1 Landscape*

An industry of M = 100 firms searches a common NK landscape. A configuration x is a binary string of length N. Each locus i has a dependency set D_i containing K other loci, drawn uniformly at random and fixed for the run, and a contribution function c_i defined over its own value and the values of its K dependencies.

For each of the 2^(K+1) input combinations, c_i is drawn independently from Uniform(0,1).

The performance of configuration x is f(x) = (1/N) Σ_i c_i(x_i, x_{D_i}).

We report *N* = 48 in the main text and replicate at $N \in \{32,64\}$. *K* runs over a five-point grid spanning nearly separable to nearly fully coupled landscapes. Note the convention: *N* counts loci; the firm count is *M* and is always 100.

### *A.2 The round*

Write $x_t^{(j)}$ for firm j's configuration at round *t* and $\mathcal{N}_1(x)$ for its single-locus neighborhood. A round proceeds:

1. **Local improvement.** If there exists $y \in \mathcal{N}_1\left(x_t^{(j)}\right)$ with $f(y) > f\left(x_t^{(j)}\right)$, the firm moves to the best such y, charging one move.
2. **Stuck.** Otherwise the firm is at a local peak, and social imitation becomes available.
3. **Source.** The firm draws a candidate from its assigned source (§A.3).
4. **Evaluation.** A firm with capacity *m* draws *m* candidates and evaluates $f(\cdot)$ at the configuration each would produce, adopting the best if it improves on the status quo. A firm with *m* = 1 evaluates one; a blind firm adopts without evaluating.
5. **Repair.** After adoption the firm resumes local improvement from wherever the adoption left it.

Every changed locus is recorded in the move counter. Social imitation is available while a firm's stage allocation remains; improving hill-climbing moves are completed and recorded even if they take the count slightly above that allocation. We set the social-learning allocation to 20 per stage with one renewal at t =

200. All arms in a comparison face the same horizon and allocation, and realized movement is reported explicitly.

Priority robustness protocol. The current policy reproduces the archived gate exactly. Under the strict-locus policies, a local-search or repair flip costs one locus, a social adoption costs its realized Hamming displacement, and every realized configuration change is admitted only if the full cost fits within the stage allocation; multi-locus moves are atomic and never partially applied. Allocations are 20 or 40 loci and renew at the disruption boundary. The uncapped policy records the same actual-locus ledger without an affordability gate and extends both stages from 200 to 400 rounds. All variants use N=48, M=100, K in {4,12,24,36,46}, 200 replications per K, the same seed architecture and disruption draws, and paired no-, partial-, and complete-disruption cells.

### *A.3 Sources*

Let $\mathcal{P}_j$ be firm j's observation neighborhood (ten firms, fixed). Write $\mathcal{C}_t$ for the published catalog at round *t* (§B).

| arm | source | adoption |
|---|---|---|
| `M1` | none | — (solitary search) |
| `M2` | the best-performing peer in the neighborhood | blind: copy a random subset of that peer's loci |
| `M2_EVAL_SUBSET(m)` | as `M2` | evaluate m subsets of that peer, adopt the best |
| `M2_EVAL_PEER(m)` | m peers from the neighborhood | evaluate one subset from each, adopt the best |
| `A_EVAL(1)` | the catalog at round t | evaluate one drawn bundle, adopt if it improves |
| `A_EVAL(m)` | the catalog at round t | evaluate m drawn bundles, adopt the best if it improves |
| `A_MATCH(m)` | the catalog at round t | **matched placebo**: adopt a bundle at random, matched to `A_EVAL(m)` on firing rate and bundle size |

`A_MATCH(m)` is the comparator for every content claim in the paper. It fires as often as `A_EVAL(m)`, takes bundles of the same size from the same catalog, and is matched on spend before any outcome is read. The difference between `A_EVAL(m)` and `A_MATCH(m)` is therefore the value of *judging*, net of the value of *being moved*.

### *A.4 Outcomes and interpretation*

The primary outcome is discovery, measured as the best configuration found. Population-mean performance and population diversity are secondary outcomes reported separately. A mean-only gain is not described as a discovery gain, particularly when it coincides with a lower best-found value or a collapse in diversity.

### *A.5 Seeding and statistics*

The landscape, the initial population, the observation network and the disruption are seeded from the tuple (run seed, *N*, *K*, replication index) only. Arm-level stochasticity adds the arm and condition index. The same landscapes therefore appear in every condition, and contrasts are paired exactly on $(K,\text{replication})$.

For a paired contrast with *n* pairs, we report $\bar{d} \pm t_{0.975,n-1}\, s_d/\sqrt{n}$. Degrees of freedom are always $n-1$; we never use a normal approximation.

## B. The abstraction operator and what it can publish

### *B.1 Definition*

Let E_t = {e_1,...,e_q}, with q at most E = 10, denote the highest-performing distinct configurations present at round t after duplicate firm configurations are collapsed. A pattern is a partial assignment P = {(i,v_i): i in I}, where I is a subset of the N loci and v_i is binary. Configuration e exhibits P if e_i = v_i for every i in I. The support of P is

$$\mathrm{supp}(P) \;=\; |\{\, j \in \mathcal{E}_t \;:\; j \text{ exhibits } P \,\}|.$$

A pattern is *closed* if no strict superset of it has the same support. The catalog is the set of closed patterns with $\mathrm{supp}(P) \geq s_{\min}$ and $|I| \leq L_{\max}$, where τ = 0.25 gives a support floor of $s_{\min} = 2$ of the 10 leading distinct configurations and $L_{\max} = \lceil N/3 \rceil$; at most *A* = 100 are retained for firms to draw from. **A bundle therefore requires only a small minority of the elite to share it**, which is why the catalog holds many overlapping entries rather than a single consensus.

Crucially, **no value is published**. The catalog is a set of partial assignments with no fitness attached. A firm learns what a bundle is worth only by evaluating $f(\cdot)$ at the configuration adopting it would produce.

*B.2 The bound*

For $S \subseteq \mathcal{E}_t$ non-empty, define the *agreement pattern*

$$A(S) \;=\; \left\{(i,v) \;:\; x_i^{(j)} = v \;\text{ for every } j \in S\right\}.$$

**Proposition B.1.** *Every closed pattern of $\mathcal{E}_t$ equals* A*(S) for some non-empty $S \subseteq \mathcal{E}_t$. Consequently the number of distinct closed patterns is at most $2^E - 1$, independent of* N.

*Proof.* Let P be closed and let $S_P = \{j \in \mathcal{E}_t\text{: } j \text{ exhibits } P\}$, so $\operatorname{supp}(P) = |S_P|$. Every configuration in $S_P$ agrees with P on I, hence $P \subseteq A(S_P)$. And every configuration in $S_P$ exhibits $A(S_P)$ by construction, so $\operatorname{supp}\big(A(S_P)\big) \geq |S_P| = \operatorname{supp}(P)$; since $A(S_P) \supseteq P$ and P is closed, $A(S_P) = P$. The map $S \mapsto A(S)$ therefore surjects the non-empty subsets of $\mathcal{E}_t$ onto the closed patterns, of which there are at most $2^E - 1$.

□

With *E* = 10 the ceiling is 1,023 patterns however large *N* is. The mined catalog is thus a bounded object whose size is governed by the elite's internal structure, not by the size of the problem.

*B.3 The raw-material constraint*

**Corollary B.2. If all E elite configurations hold the same configuration x, every nonempty subset of elites yields the same agreement pattern. The catalog therefore contains exactly one pattern and offers nothing to choose among.**

This is the formal statement of §2.2. What a community can publish is bounded by what its leading members do *not* have in common. In a homogeneous industry the intersection lattice collapses, the catalog degenerates to a single item, and the firm's capacity to weigh *m* candidates has nothing to weigh. It is why the return to that capacity in our most homogeneous condition is +0.0004 and in our most diverse is +0.0091.

The corollary also explains why a *quality filter* cannot repair a thin catalog. Filtering selects a subset of $A(\cdot)$; it cannot create patterns that the elite's agreement structure does not contain.

## C. Why option diversity and not option quality

### *C.1 The general statement*

Let $X_1, \ldots, X_m$ be exchangeable random variables representing the fit of $m$ candidate bundles to a given firm's configuration, and define the value of the capacity to weigh $m$ candidates as

$$V(m) \; = \; \mathbb{E}\left[\max_{1 \le i \le m} X_i\right] \; - \; \mathbb{E}[X_1].$$

**Proposition C.1 (location invariance, scale homogeneity). Let Y_i = a + bX_i with b > 0. Then V_Y(m) = bV_X(m) for all m. Thus V is invariant to a pure shift in candidate quality and homogeneous of degree one in scale.**

*Proof. Because b > 0, max_i(a + bX_i) = a + b max_i(X_i). Taking expectations and subtracting the value of one draw gives the result.* ■

The proposition supplies a conditional benchmark. If a catalog manipulation shifts every candidate by a common amount while leaving dispersion and dependence unchanged, it does not change the value of choosing among candidates. Manipulations that also change dispersion, dependence, or the accept-or-reject margin need not satisfy this invariance.

### *C.2 The Gaussian case, and a testable shape*

If $X_i \sim \mathcal{N}(\mu, \sigma^2)$ independently, then $\mathbb{E}[\max_i X_i] = \mu + \sigma a_m$ where $a_m = \mathbb{E}[\max_{i \le m} Z_i]$ for standard normal $Z_i$, so

$$V(m) \; = \; \sigma \, a_m, \qquad a_m = m \int_{-\infty}^{\infty} z \; \phi(z) \, \Phi(z)^{m-1} \, dz.$$

Under the independent Gaussian benchmark, the constants imply that the ladder in m is linear in the expected-normal-maximum term a_m, with slope σ. The benchmark therefore offers a descriptive shape against which to compare the simulated ladder; it is not a distributional assumption of the model.

### *C.3 The fit*

Regressing the measured content premium at $N = 48$ on $a_m$ across the six values of $m$:

| catalog | intercept | slope (σ) | $R^2$ |
|---|---|---|---|
| performance-screened | 0.00480 | 0.00966 | 0.9969 |
| top-performer source | 0.00464 | 0.01053 | 0.9970 |
| random source | 0.00816 | 0.00939 | 0.9989 |

Three things are worth drawing out.

**The Gaussian benchmark describes the observed shape closely: the regression $R^2$ exceeds 0.997 in each catalog condition. This high fit is useful as a compact summary of the ladder, but it does not validate Gaussian, independence, or location-scale assumptions outside these simulated cells.**

**The fitted slope is similar across catalog conditions—0.0094 to 0.0105, a range of 0.0011. Within this benchmark, the slope summarizes the scale on which candidate bundles differ in fit. The result is consistent with, rather than a direct measurement of, Proposition C.1's location-scale mechanism.**

**The intercept captures the single accept-or-reject margin available at m = 1. It is 70% larger for the random-source catalog (0.0082 versus 0.0046–0.0048), consistent with that catalog containing more candidates that a firm is right to reject. This decomposition is descriptive: the random-source catalog does not appear to offer a wider option set, but it offers more to refuse.**

**Why the population's diversity governs σ.** By Corollary B.2, the candidate bundles are agreement patterns of the elite. As the elite becomes more alike, those patterns converge on a single item and the differences in fit across drawn candidates shrink; $\sigma \to 0$ and, by C.2, $V(m) \to 0$ for every $m$. The measured ladder falls from +0.0091 to +0.0004 across our diversity range, which is what that limit predicts.

## D. The disruption operator

### D.1 Definition

At round $T_d = 200$ a set $D \subseteq \{1, \dots, N\}$ with $|D| = \lfloor \varphi N \rceil$ is drawn uniformly. For each $i \in D$:

- **weights variant:** the contribution table $c_i$ is re-drawn independently from the same distribution;
- **weights-and-structure variant (reported throughout):** the dependency set $B_i$ is re-drawn uniformly *and* the table $c_i$ is re-drawn on the new arguments.

No firm's configuration is altered; what changes is what configurations are worth. This is a shift in the payoff environment of the kind Posen and Levinthal (2012) study, with the architectural component that Henderson and Clark (1990) emphasize added in the second variant.

### D.2 The correlation between the pre- and post-disruption landscape

**Proposition D.1.** *Let f and f' denote performance before and after the disruption. For a configuration x drawn independently of the tables,*

$$\mathrm{Corr}\big(f(x),\, f'(x)\big) \;=\; 1 - \varphi.$$

*Proof. The pre- and post-disruption performance functions share the contribution terms for loci outside D; contributions for loci in D are independent redraws. With independent locus contributions and common variance σ_c², the two functions share exactly N(1 − φ) terms, so*

$$\mathrm{Cov}(f, f') = \frac{1}{N^2} \sum_{i \notin D} \mathrm{Var}\,(c_i) = \frac{(1-\varphi)\,\sigma_c^2}{N}, \qquad \mathrm{Var}(f) = \mathrm{Var}(f') = \frac{\sigma_c^2}{N},$$

and the covariance-to-variance ratio is $1 - \varphi$. ■

We derived this before reading any output from the disruption module and it is confirmed to four decimal places at every φ we run, in both variants. It is the one piece of the implementation validated against an analytic result rather than against itself, and we report it as such.

### D.3 Catalog inertia

The parameter $\lambda \in [0,1]$ is the fraction of the published catalog carried over unrevised after the disruption, retaining its *recorded* support — that is, the support it had before the world changed. Setting $\lambda = 0$ re-mines the catalog from scratch each round, which is the assumption we use everywhere except §5.2, so that inertia is never confounded with the disruption. Setting λ > 0 models the realistic case in which a repertoire is revised more slowly than the world moves.

## E. Estimation of the crossing

### *E.1 The estimand*

Let $d_c$ be the realized pre-disruption population diversity of cell c and $y_c$ the paired effect of the disruption on discovery in that cell, with standard error $s_c$. Fit

$$y_c \;=\; \alpha + \beta\, d_c + \varepsilon_c, \qquad w_c = s_c^{-2},$$

by weighted least squares across the seven cells. The crossing is the ratio

$$d^* \;=\; -\hat{\alpha}/\hat{\beta}.$$

### *E.2 Why the cell means and not the replications*

A per-replication regression has 7,000 observations behind it and looks more authoritative. It is attenuated. Within a cell, realized diversity varies mostly as sampling noise rather than as the designed contrast, so regressing on it is a classical errors-in-variables problem and shrinks $\hat{\beta}$ toward zero. The seven cell means are the designed contrast; each is measured to about 0.008 in diversity and 0.0005 in effect, and weighting by precision uses the design as it was built.

| estimator | φ = 1.00 | φ = 0.25 |
|---|---|---|
| **cell-mean WLS (primary)** | **0.6759** | **0.6265** |
| per-replication OLS | 0.6620 | 0.5803 |
| bracketing interpolation | 0.7064 | 0.6470 |

The three span 0.662 to 0.706 at φ = 1.00, narrower than the primary interval itself, so the choice of estimator does not drive the conclusion. The reportable statement is “about 0.68”.

### *E.3 Fieller intervals*

A ratio of estimated coefficients has no symmetric confidence interval, and a delta-method band understates it when $\hat{\beta}$ is not overwhelmingly precise. We invert the test instead: the 1 - γ confidence set for $d^*$ is

$$\left\{ d \,:\, \left(\hat{\alpha} + \hat{\beta} d\right)^2 \;\le\; t^2_{1-\gamma/2,\,\nu} \left(\hat{V}_{\alpha\alpha} + 2d\hat{V}_{\alpha\beta} + d^2 \hat{V}_{\beta\beta}\right) \right\},$$

a quadratic inequality in *d* whose root interval is reported. At φ = 1.00 this gives [0.6263, 0.7202]; at φ = 0.25, [0.5793, 0.6649].

### *E.4 Monotonicity and uniqueness*

The response is monotone across all seven levels at both magnitudes, with zero significant sign changes: the effect passes through zero once and does not return. A non-monotone response or a second crossing would complicate the two-process account in Section 2.4; neither appears in these designed cells.

## F. Supporting tables

### *F.1 The seven-cell design in levels*

*N = 48, weights-and-structure disruption, λ = 0, top-performer-source catalog, arm A_EVAL(10), 400 rounds, two nominal social-learning allocations, n = 1,000 pairs per cell. Diversity is the share of distinct configurations in the φ = 0 control at round 200.*

| pre-disruption diversity | φ = 0 | φ = 0.25 | φ = 1.00 |
|---|---|---|---|
| 0.3994 | 0.71078 | 0.71361 | 0.71605 |
| 0.5729 | 0.71674 | 0.71855 | 0.72025 |
| 0.6570 | 0.72052 | 0.72028 | 0.72225 |
| 0.7163 | 0.72354 | 0.72192 | 0.72319 |
| 0.7583 | 0.72529 | 0.72348 | 0.72358 |
| 0.7934 | 0.72782 | 0.72488 | 0.72421 |
| 0.9752 | 0.74020 | 0.73396 | 0.73173 |

Paired contrasts, most diverse minus most homogeneous: +0.02942 ± 0.00109 at φ = 0; +0.02035 ± 0.00101 at φ = 0.25; +0.01568 ± 0.00096 at φ = 1.00. Diverse-and-fully-disrupted minus homogeneous-and-never-disrupted: +0.02095 ± 0.00109.

### *F.2 Improving moves completed and the nominal allocation*

| pre-disruption diversity | control | φ = 0.25 | φ = 1.00 |
|---|---|---|---|
| 0.3994 | 12.69 | 18.73 | 23.78 |

| pre-disruption diversity | control | φ = 0.25 | φ = 1.00 |
|---|---|---|---|
| 0.5729 | 16.04 | 22.19 | 26.88 |
| 0.6570 | 18.20 | 24.03 | 28.56 |
| 0.7163 | 20.12 | 25.74 | 29.96 |
| 0.7583 | 21.61 | 27.06 | 30.87 |
| 0.7934 | 22.85 | 27.90 | 31.89 |
| 0.9752 | 37.08 | 38.25 | 40.27 |

*F.3 The two channels, correlated against the gain in discovery*

Within-landscape correlations, averaged across the five landscape settings; the three least diverse conditions.

| condition | scattering channel | reopened-opportunity diagnostic |
|---|---|---|
| r = 2, φ = 0.25 | −0.050 | +0.110 |
| r = 2, φ = 1.00 | −0.053 | +0.106 |
| r = 4, φ = 0.25 | −0.131 | +0.124 |
| r = 4, φ = 1.00 | −0.074 | +0.060 |
| r = 6, φ = 0.25 | −0.114 | +0.108 |
| r = 6, φ = 1.00 | −0.018 | +0.038 |

Zero of six positive for scattering; six of six positive for the gradient, significant in four.

*F.4 The flip at three problem sizes*

| N | homogeneous: diversity, effect | diverse: diversity, effect | n |
|---|---|---|---|
| 32 | 0.2156, +0.00328 ± 0.00091 | 0.7803, -0.00864 ± 0.00076 | 2500 |
| 48 | 0.3994, +0.00527 ± 0.00105 | 0.9752, -0.00848 ± 0.00090 | 1000 |
| 64 | 0.4832, +0.00474 ± 0.00130 | 0.9999, -0.00848 ± 0.00117 | 500 |

All six significant. Note the realized diversity at the same nominal setting: 0.2156, 0.3994, 0.4832 — a range of 0.27, which is why the threshold is stated in diversity.

*F.5 What a disruption does to the catalog*

| φ | practices retaining support | catalog diversity at T+1 | as % of pre-disruption |
|---|---|---|---|
| 0 | 99.9% | 0.04309 | 100% |
| 0.10 | 99.2% | 0.04161 | 96.6% |
| 0.25 | 97.9% | 0.03855 | 89.5% |
| 0.50 | 96.2% | 0.03203 | 74.3% |

| φ | practices retaining support | catalog diversity at T+1 | as % of pre-disruption |
|---|---|---|---|
| 0.75 | 94.7% | 0.02712 | 62.9% |
| 1.00 | 93.4% | 0.02505 | 58.1% |

*F.6 The cost of a community norm*

Forcing intensity h; N = 48, top-performer-source catalog, static landscape, n = 1000.

| h | discovery vs free choice | distinct configurations | mean pairwise distance | return to judging |
|---|---|---|---|---|
| 0 | — | 0.922 | 18.14 | +0.00952 ± 0.00081 |
| 0.25 | -0.00180 ± 0.00070 | 0.938 | 18.38 | +0.00777 ± 0.00076 |
| 0.50 | -0.00306 ± 0.00074 | 0.955 | 18.83 | +0.00673 ± 0.00076 |
| 1.00 | -0.00988 ± 0.00078 | 1.000 | 22.63 | 0.00000 |

*F.7 Robustness of the disruption implementation*

The two disruption variants — re-drawing payoffs alone, and re-drawing payoffs while rewiring dependencies — are indistinguishable at every magnitude on every measure we report. The main text uses the weights-and-structure variant throughout. As one instance, the share of practices losing support at φ = 1.00 is 0.0665 under weights and 0.0664 under weights-and-structure; catalog diversity at $T + 1$ is 0.02527 and 0.02505.

The collapse in catalog diversity and in the return to judgment replicates at both other problem sizes: catalog diversity at φ = 1.00 falls to 0.0271 at $N = 32$ and 0.0224 at $N = 64$, with the return to judging falling to zero in both.

*F.8 Registered predictions*

Predictions were written into the run drivers before execution and never revised; the driver headers are preserved in the replication package. Where a registered prediction ran against a claim held in the literature, the result is reported in the main text as a result. Registered predictions concerning internal conjectures are not reported as findings; two are reported because they discriminate between competing mechanisms and the discrimination is load-bearing: the scattering channel (§5.3), which the registered prediction expected to hold and which the data contradict, and the direction of the crossing in φ (§5.4).

## G. Additional boundary conditions and implementation details

### *G.1 Diversity measures and the crossing*

The simulated crossing is expressed in model units and should not be exported as a universal threshold. Distinct share depends on the fixed population of M = 100 firms and approaches its maximum as nearly every firm becomes unique. Normalized pairwise Hamming distance and locus entropy avoid that ceiling, but measure different properties on different scales and consequently yield different fitted crossings. We report each measure separately and estimate a crossing only when the observed conditions lie on both sides of zero. Empirical work should trace diversity among leading sources and dispersion in candidate fit without treating any simulated crossing as universal.

### *G.2 Limits of the mechanism comparison*

Matched random-movement conditions separate locally selected catalogue content from displacement, but they do not complete causal mediation. Across search-limit designs, the advantage of locally selected content rises at low diversity and falls at high diversity. Displacement alone is positive at low diversity in three designs but not under the hard 20-decision cap. A stronger factorial experiment would independently vary four repertoire conditions—none, retained pre-disruption content, content re-mined from current leaders, and content explicitly validated after disruption—and cross each with the presence or absence of situated judgment. The main text therefore treats the present comparisons as mechanism evidence rather than complete identification.

### *G.3 Search budgets and observation horizons*

The archived implementation limits social imitation but allows an improving one-decision local move to pass the nominal allowance. Robustness analyses impose hard 20- and 40-decision stage limits on every realized change and also test a 400+400-round horizon without a decision-change limit. Even under the longer horizon, a material share of runs is still improving when observation stops, so we do not describe the endpoint as convergence. These checks address budget headroom and simulation length within the same search architecture; they do not establish robustness to every landscape, network, or behavioral rule.

### *G.4 Landscape, population, and network scope*

The model uses binary NK landscapes, M = 100 firms, one local-search rule, and a fixed observation network. Results replicate at N = 32, 48, and 64 and across five levels of interdependence. Other landscape structures, population sizes, dynamic networks, and search rules may shift the crossing or remove it. These are scope conditions on the quantitative result, not changes to the proposed two-process logic.

### *G.5 Abstraction rule and source diversity*

The model compresses distributed abstraction into a transparent recurrence rule applied to up to ten leading distinct configurations. Because those sources are selected to be distinct, the share that is distinct is always one and cannot discriminate among conditions; only their distance from one another and their decision-by-decision diversity are informative. Institutions that test causal performance, deliberately represent minority experience, or preserve contextual information may be less vulnerable to the popularity–usefulness gap. The gap is therefore a boundary condition of recurrence-based curation, not a claim about every knowledge institution.

### *G.6 Analytic and replication safeguards*

The relationship between disruption magnitude and similarity of the pre- and post-disruption landscapes was derived analytically and then confirmed computationally. Focal mechanism predictions were recorded before the simulations were interpreted, and the complete code and replication materials accompany the paper. These safeguards support implementation validity but do not remove the model's substantive boundary conditions.